\documentclass[aps,pre,twocolumn,showpacs,amsmath,amssymb]{revtex4-2}

\usepackage[hidelinks]{hyperref}
\usepackage{graphicx}
\usepackage{subfig}

\usepackage[export]{adjustbox}
\usepackage{color}
\usepackage[utf8]{inputenc}
\usepackage[T1]{fontenc}
\usepackage{diagbox}
\usepackage{multirow}
\usepackage{bm}
\usepackage{physics}

\begin{document}

\title{Imbibition dynamics under dissolution-driven fluid thickening}

\author{Hubert W. Almeida$^1$}
\author{Andr\'{e} F.~V. Matias$^{1,2}$}
\author{Rodrigo C.~V. Coelho$^{1,3}$}
\author{Nuno A.~M. Ara\'{u}jo$^{1,4}$}
\email{nmaraujo@ciencias.ulisboa.pt}
\affiliation{$^1$Centro de F\'{i}sica Te\'{o}rica e Computacional, Faculdade de Ci\^{e}ncias, Universidade de Lisboa, 1749--016 Lisboa, Portugal}
\affiliation{$^2$Soft Condensed Matter \& Biophysics, Debye Institute for Nanomaterials Science, Utrecht University, Princetonplein 1, 3584 CC Utrecht, The Netherlands}
\affiliation{$^3$Centro Brasileiro de Pesquisas Físicas, Rua Xavier Sigaud 150, 22290-180 Rio de Janeiro, Brazil}
\affiliation{$^4$Departamento de F\'{i}sica, Faculdade de Ci\^{e}ncias, Universidade de Lisboa, 1749--016 Lisboa, Portugal}

\begin{abstract}
The imbibition of milk into a biscuit provides an example of a broader class of processes in which a liquid invades a porous material, dissolving components of the solid and thereby modifying its own rheology. This coupling between imbibition, dissolution, and evolving fluid rheology also occurs in food processing, filtration, and reactive porous media. While the resulting changes in fluid viscosity are known to affect imbibition, a quantitative description of how this evolving rheology determines the imbibition dynamics is still lacking. Here, we establish such a description for dissolution-driven fluid thickening. For a capillary tube in the well-mixed limit, we derive a generalized Washburn law showing that the evolving viscosity can be incorporated through a rescaled time given by the accumulated inverse viscosity. This predicts a collapse of the imbibition dynamics for different dissolution rates, which we confirm numerically using coupled free-surface and solute-transport simulations. When diffusion is slower, concentration and viscosity become spatially heterogeneous, resulting in a weaker retardation of the advancing liquid than predicted by the well-mixed theory. We then consider disordered porous media and quantify how dissolution-driven thickening and variations in pore geometry affect the dynamics. Our results extend the classical Washburn description to imbibition with evolving fluid properties and provide a quantitative basis for understanding fluid uptake when interactions with the solid progressively modify the rheology of the invading liquid.
\end{abstract}

\maketitle

\section{Introduction}
Dunking a biscuit in milk causes the liquid to be absorbed while the biscuit progressively softens. This familiar observation illustrates a broader class of processes in which capillary-driven imbibition is coupled to changes in both the invading liquid and the material it penetrates. As the liquid penetrates the biscuit, components of the solid dissolve in it, modifying the properties of the liquid, while swelling and other structural changes can alter the pore space through which it flows~\cite{Fisher1999}. Similar coupled processes arise in a wide range of natural and technological systems, including soil hydration, oil recovery, filtration, paper wetting, and food processing~\cite{CAI2022102654, SAGUY200537}. In many of these systems, fluid invasion occurs in porous structures whose geometry and physicochemical properties evolve over time due to swelling, dissolution, deposition, erosion, or clogging, while the invading fluid may simultaneously undergo substantial changes in transport properties such as viscosity and diffusivity as a consequence of solute release or chemical reactions. The resulting feedback between fluid invasion, solute transport, and material properties can significantly modify fluid uptake and transport~\cite{PhysRevFluids.8.094002}.

Most studies of such feedback mechanisms have focused on changes in the porous medium and on how the resulting evolution of the pore geometry modifies the flow. The importance of time-dependent porous structures has been highlighted in studies of flow through evolving media~\cite{https://doi.org/10.1029/2019JB017813, MATIAS2021101360, Jager2017, Matias10.1063/5.0176273, CoelhoChannelization, ELLERO2019181, Kudrolli2016}. In particular, swelling and erosion can strongly affect permeability and transport in deformable and reactive porous materials~\cite{Matias10.1063/5.0176273, ELLERO2019181, Aksu2015, Mateus2007}, while dynamically coupled erosion, deposition, and flow can lead to channelization and discontinuous transitions toward localized flow patterns~\cite{CoelhoChannelization}. Related effects occur in granular and powder-based systems relevant to the food industry, where moisture absorption and plasticization modify inter-particle viscosity and promote caking~\cite{Simoes2022,Braz2022,CardenasBarrantes26}. In these examples, fluid--solid interactions affect transport primarily by modifying the medium through which the fluid moves. The complementary feedback arises when the material released by the solid changes the rheology of the invading fluid. Recent pore-scale studies of CO$_2$ dissolution-induced viscosity reduction in the displaced oil phase during injection processes have begun to address such dynamically evolving fluid properties~\cite{Zhang2026}. However, to the best of our knowledge, quantitative descriptions of how solute uptake from a porous matrix and the resulting continuous evolution of the fluid viscosity determine spontaneous imbibition dynamics are still lacking.

From a hydrodynamic perspective, capillary imbibition in rigid porous media is commonly described by Washburn's law, which predicts a square-root-in-time growth of the penetration height under the assumptions of constant viscosity and static pore geometry~\cite{PhysRev.17.273}. Deviations from this classical behavior have been reported in systems involving complex wetting dynamics, multiphase interactions, and non-Newtonian effects~\cite{HU2026139583, PhysRevApplied.14.054009, Zhou2024, 10.1063/5.0228835}. Wetting properties and contact-angle dynamics significantly influence capillary rise and multiphase flow~\cite{10.1063/5.0228835, 10.1063/5.0302948}. Dissolution-driven fluid thickening introduces a distinct departure from the classical Washburn picture: the viscosity evolves during imbibition because the invading liquid progressively takes up material from the solid. The relevant question is therefore not simply whether an increase in viscosity slows imbibition, but how the time-dependent rheology determines the invasion dynamics and under which conditions a generalized description of capillary imbibition can be obtained.

In this work, we develop such a quantitative description by combining an analytical treatment of capillary imbibition with numerical simulations that resolve the coupled evolution of the liquid interface, solute concentration, and fluid viscosity. In the well-mixed limit, we derive a generalized Washburn law in which the evolving viscosity enters through a rescaled time determined by the accumulated inverse viscosity. This description predicts a collapse of the imbibition dynamics for different dissolution rates, which we confirm numerically. We then relax the well-mixed assumption and determine how finite solute diffusion and the resulting spatial heterogeneity of the viscosity modify the capillary dynamics. Finally, we extend the analysis to disordered porous media, where the invasion becomes heterogeneous and intermittent, and investigate separately the effects of dissolution-driven fluid thickening and pore geometry. Beyond the well-mixed limit, resolving the coupling between interface motion, solute transport, and the spatially evolving viscosity field requires a numerical treatment; here, this is achieved using a free-surface lattice Boltzmann method coupled to an advection-diffusion solver.

The remainder of this paper is organized as follows. In Sec.~\ref{sec:2}, we introduce the model and develop a theoretical description of capillary imbibition under dissolution-driven fluid thickening, deriving a generalized Washburn law in the well-mixed limit. In Sec.~\ref{sec:3}, we use numerical simulations to first test this description and examine the effects of finite solute diffusion in capillary tubes (Sec.~\ref{sec:3-Capillary}), before extending the analysis to disordered porous media (Sec.~\ref{sec:3-Porous_media}). Details of the numerical methods are provided in Appendix~\ref{sec:SI-Methods}. Finally, in Sec.~\ref{sec:4}, we summarize our main findings and discuss possible extensions.

\section{Model and theoretical description} \label{sec:2}
We first consider the idealized case of a liquid imbibing a cylindrical capillary tube whose walls release a soluble material into the liquid. As the liquid advances, the dissolved material changes its viscosity, which in turn affects the imbibition dynamics. This simple geometry allows us to establish a quantitative relation between the evolving fluid viscosity and the motion of the liquid interface.

Neglecting gravity, the motion of the liquid is driven by the capillary pressure across the meniscus. For a cylindrical tube of radius $r$ and equilibrium contact angle $\theta<90^\circ$, the resulting pressure difference drives the liquid into the tube. For a liquid of constant dynamic viscosity $\eta$, the height $h(t)$ of the advancing meniscus is described by Washburn's law~\cite{PhysRev.17.273}:
\begin{equation}
    h^2(t) = \frac{\gamma r \cos\theta}{2 \eta} t \,,
    \label{eq:washburn_law}
\end{equation}
where $\gamma$ is the surface tension.

We now consider the effect of solute released from the walls. The evolution of the solute concentration is governed by the competition between transport by the advancing liquid, diffusion within the liquid, and transfer of solute from the solid surface. The relative importance of the advective and diffusive transport is characterized by the Péclet number,
\begin{equation}
    \mathrm{Pe} = \frac{U L}{D},
\end{equation}
where $U$ is a characteristic velocity (here is the initial meniscus velocity $\dd h / \dd t$), $L$ a characteristic length scale (here is the radius $r$), and $D$ the diffusion coefficient of the solute.

\begin{figure*}[t]
    \includegraphics[width=\linewidth]{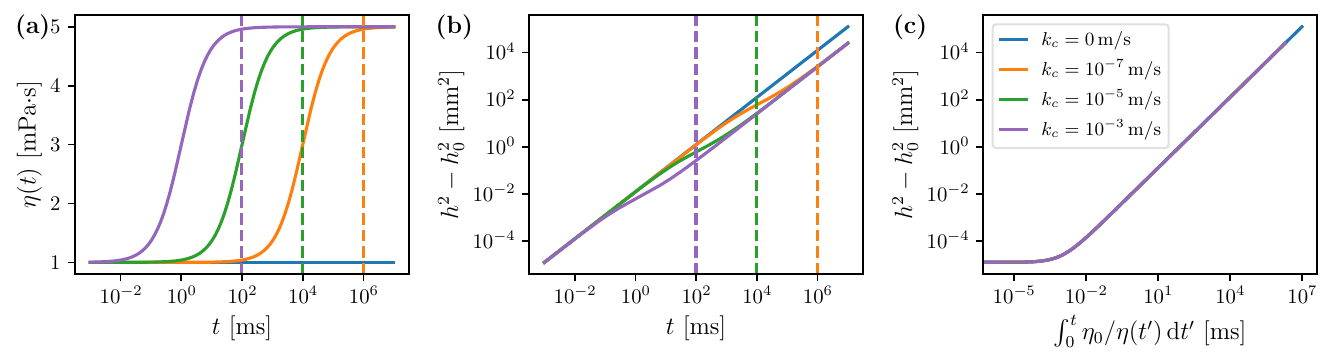}
    
    \caption{Analytical results for capillary imbibition in the well-mixed limit for different mass-transfer rate coefficients $k_c$.  (a) Evolution of the dynamic viscosity $\eta(t)$ obtained from the linear concentration increase and the concentration-dependent viscosity given by Eq.~\eqref{eq:viscosity}. Increasing $k_c$ accelerates the approach to the asymptotic viscosity $\eta_{\mathrm{sat}}=\eta_0(1+\beta)$, with a characteristic crossover timescale $\tau_{\mathrm{sat}}\sim k_c^{-1}$ (vertical dashed lines). (b) Corresponding evolution of the squared meniscus displacement, $h^2-h_0^2$, as given by Eq.~\eqref{eq:varying_viscosity_height}. At early times, the dynamics follow the classical Washburn law with viscosity $\eta_0$; as the viscosity increases, imbibition slows and progressively crosses over to Washburn-like behavior governed by $\eta_{\mathrm{sat}}$. (c) The same imbibition curves plotted as a function of the rescaled time $\tilde{t}=\int_0^t[\eta_0/\eta(t')]\,\dd t'$. As predicted by the generalized Washburn law, Eq.~\eqref{eq:general_solution}, the curves for different $k_c$ collapse onto the classical Washburn form, $h^2-h_0^2=(\gamma r\cos\theta/2\eta_0)\tilde{t}$.}
    \label{fig:collapse}
\end{figure*}

We first consider the well-mixed limit, $\mathrm{Pe}\ll 1$, in which diffusion is sufficiently fast to maintain an approximately uniform solute concentration throughout the liquid column, such that $C(\mathbf{x},t)\simeq C(t)$. Assuming that the hydrodynamic response remains quasistatic as the viscosity evolves, Washburn's equation can be written in terms of the instantaneous viscosity $\eta[C(t)]$, while the evolution of the solute concentration follows from a mass balance. The coupled dynamics are then described by
\begin{equation}
    \begin{cases}
        \displaystyle \dv{h(t)}{t} = \frac{\gamma r \cos \theta}{4 \eta[C(t)] h(t)} \vspace{0.5em} \\
        \displaystyle \dv{C(t)}{t} = \frac{k_c C_s A}{V}
        \approx 2 \frac{k_c C_s}{r}
    \end{cases} \, ,
    \label{eq:systemeq}
\end{equation}
where $k_c$ is the mass-transfer coefficient, $C_s$ is the solute concentration at the solid surface, $A$ is the surface area of the capillary wall in contact with the liquid, and $V$ is the volume of liquid inside the tube. The dissolution rate follows from the source term in the advection--diffusion equation, as detailed in Appendix~\ref{sec:SI-Dissolution}.

Taking into account the initial condition $C(0)=0$, the solute concentration increases linearly with time,
\begin{equation}
    C(t) = 2 \frac{k_c C_s}{r} t \, .
\end{equation}
The viscosity therefore becomes time dependent through the relation $\eta(t)=\eta[C(t)]$, without requiring at this stage any specific functional dependence of viscosity on concentration. From Eq.~\ref{eq:systemeq}, the equation governing the height of the meniscus can then be written as
\begin{equation}
    \dv{}{t}{\left(\frac{h^2(t)}{2}\right)}
    = \frac{\gamma r \cos \theta}{4}\frac{1}{\eta(t)} \, .
\end{equation}
Integrating in time yields a generalized expression for the capillary rise of a liquid with an evolving viscosity,
\begin{equation}
    h^2(t) =
        \frac{\gamma r \cos \theta}{2}
        \int_0^t \frac{1}{\eta(t')} \dd t'
        + h_0^2
     \, ,
    \qquad h_0 = h(0) \, .
    \label{eq:general_solution}
\end{equation}
Equation~\eqref{eq:general_solution} generalizes Washburn's law to a time-dependent viscosity. The imbibition dynamics are governed by the accumulated inverse viscosity, $\int_0^t \eta^{-1}(t')\,\dd t'$, irrespective of the specific functional dependence of viscosity on concentration.

To obtain an explicit solution and later perform the numerical analysis, we now specify the dependence of viscosity on solute concentration. We model the dynamic viscosity as
\begin{equation}
    \eta[C(t)] = \eta_0 \left(1 + \beta \frac{C(t)}{C(t) + \delta}\right),
    \label{eq:viscosity}
\end{equation}
where $\eta_0$ is the dynamic viscosity of the pure liquid and $\beta$ and $\delta$ control, respectively, the magnitude and concentration dependence of the increase in viscosity. In particular, $\beta$ determines the maximum relative increase in viscosity,
$\eta(C\to\infty)=\eta_0(1+\beta)$, while $\delta$ sets the concentration at which half of this increase is attained,
$\eta(C=\delta)=\eta_0(1+\beta/2)$. In this work, we set $\beta=4$ and $\delta=40\,$kg/m$^3$. This functional form captures two limiting behaviors relevant to the present model. At low solute concentrations, $C\ll\delta$, the viscosity increases approximately linearly with concentration, whereas for $C\gg\delta$ it approaches the finite value $\eta_0(1+\beta)$. The latter prevents the viscosity from increasing without bound, while we assume that the solute concentration remains well below its saturation value, such that dissolution can continue as long as the liquid remains in contact with the soluble material.

Substituting Eq.~\eqref{eq:viscosity} and the linear concentration
evolution into Eq.~\eqref{eq:general_solution} gives an explicit expression
for the meniscus height. Defining
${a \equiv \gamma r \cos\theta/(2\eta_0)}$ and
${d \equiv \delta r/(2k_c C_s)}$, we obtain
\begin{align}
    h(t) = &\frac{1}{\beta + 1}
    \biggl\{
    a (\beta + 1)t
    + h_0^2(\beta+1)^2
    \nonumber \\
    &+ a \beta d
    \log\left[
    1+\frac{(\beta+1)}{d}t
    \right]
    \biggr\}^{1/2}.
    \label{eq:varying_viscosity_height}
\end{align}
In the limits of very slow or very fast dissolution,
${k_c \ll \delta r/C_s}$ and ${k_c \gg \delta r/C_s}$, respectively,
Eq.~\eqref{eq:varying_viscosity_height} reduces to Washburn-like
behavior (see Appendix~\ref{sec:SI-Washburn}). For slow dissolution,
the concentration changes negligibly over the relevant imbibition
timescale and $\eta\simeq\eta_0$. In the opposite limit, the viscosity
rapidly approaches its asymptotic value
$\eta_{\mathrm{sat}}=\eta_0(1+\beta)$, after which the dynamics are
again described by Washburn's law, but with $\eta_{\mathrm{sat}}$.

For intermediate dissolution rates, the evolving viscosity produces
a crossover between these two regimes. As shown in
Fig.~\ref{fig:collapse}a, increasing $k_c$ causes the viscosity to
approach $\eta_{\mathrm{sat}}$ more rapidly, with a characteristic
timescale $\tau_{\mathrm{sat}}\sim k_c^{-1}$. The corresponding
imbibition dynamics are shown in Fig.~\ref{fig:collapse}b.
At early times, before the viscosity has changed appreciably, the
liquid follows the classical Washburn scaling
$h^2-h_0^2\sim t/\eta_0$. As the viscosity increases, imbibition
progressively slows. At sufficiently long times, once the viscosity
has approached $\eta_{\mathrm{sat}}$, Washburn-like scaling is
recovered with the larger viscosity,
$h^2\sim t/\eta_{\mathrm{sat}}$.

\begin{figure}[t]
    \includegraphics[width=\linewidth]{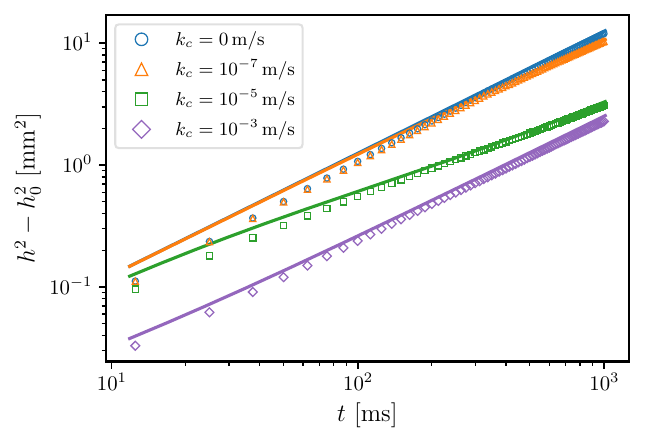}
    
    \caption{Numerical validation of the analytical description for capillary imbibition in the well-mixed regime. The height of the meniscus $h(t)$ is shown for different mass-transfer coefficients $k_c$, corresponding to different rates of solute release and viscosity increase. The simulations are performed at $\mathrm{Pe}=10^{-1}$, such that diffusion maintains an approximately uniform solute concentration within the liquid. Numerical results are compared with the analytical prediction of Eq.~\eqref{eq:varying_viscosity_height} (solid curves). The case $k_c=0$ corresponds to the classical Washburn limit with constant viscosity, while increasing $k_c$ leads to a progressively stronger retardation of the imbibition dynamics. The agreement between simulations and theory confirms the analytical description in the well-mixed limit.
    \label{fig:variable_viscosity}}
\end{figure}

The crossover can be understood directly from the general solution,
Eq.~\eqref{eq:general_solution}. Introducing the rescaled time
\begin{equation}
    \tilde{t}(t)
    \equiv
    \int_0^t \frac{\eta_0}{\eta(t')}\,\dd t',
\end{equation}
the generalized Washburn's law becomes
\begin{equation}
    h^2(t)
    =
    \frac{\gamma r\cos\theta}{2\eta_0}\tilde{t}(t)
    +h_0^2 \ \ .
\end{equation}
Thus, the effect of the evolving viscosity can be absorbed entirely
into a rescaling of time in the well-mixed limit. As shown in
Fig.~\ref{fig:collapse}c, the imbibition curves obtained for
different dissolution rates collapse onto the classical Washburn
form when plotted as a function of $\tilde{t}$.

\begin{figure}[t]
    \centering
    \includegraphics[width=\linewidth]{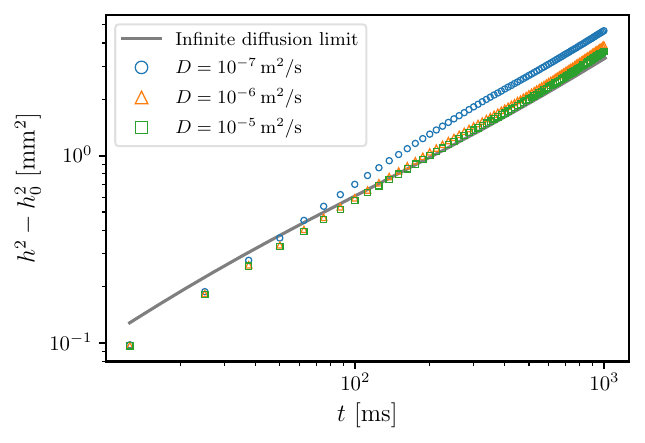}
    \caption{Effect of solute diffusion on capillary imbibition for a fixed mass-transfer coefficient $k_c=10^{-5}\,\mathrm{m/s}$. The height of the meniscus $h(t)$ is shown for diffusion coefficients $D=\{10^{-7},10^{-6},10^{-5}\}\,\mathrm{m^2/s}$, corresponding to $\mathrm{Pe}=\{10,1,10^{-1}\}$, respectively. The analytical prediction of Eq.~\eqref{eq:varying_viscosity_height}, obtained in the well-mixed limit, is shown for comparison. As $D$ increases, concentration gradients are progressively suppressed and the numerical results approach the analytical prediction. For slower diffusion, the concentration and viscosity fields become spatially heterogeneous, resulting in weaker retardation of the advancing meniscus than predicted by the well-mixed theory.}
    \label{fig:diffusion}
\end{figure}

\section{Numerical results} \label{sec:3}
We now use numerical simulations to test the theoretical description developed in Sec.~\ref{sec:2} and to investigate regimes beyond the well-mixed approximation. We first consider imbibition in a cylindrical capillary, which allows a direct comparison with the analytical results, before extending the analysis to disordered porous media.

\begin{figure*}[t]
    \includegraphics[width=\linewidth]{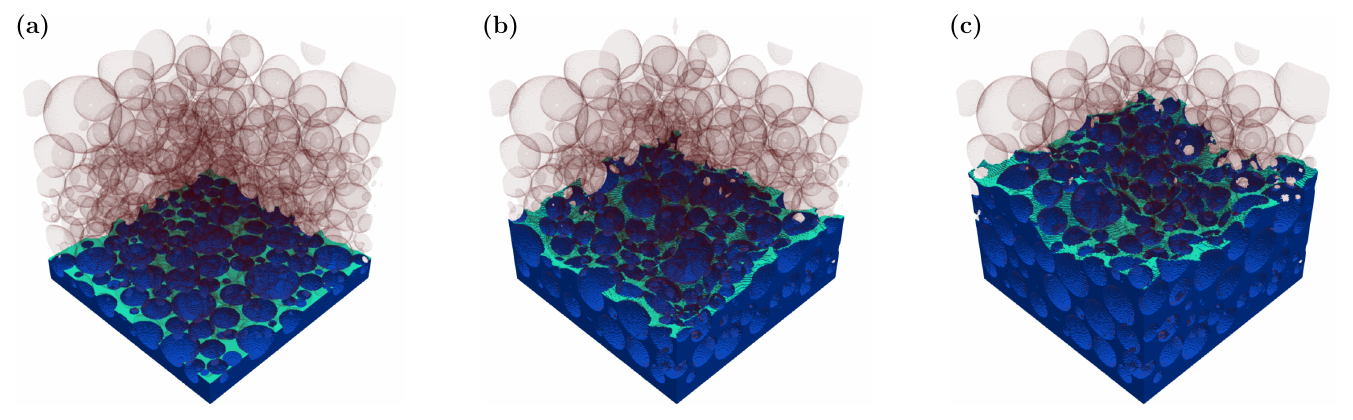}
    
    \caption{Example of capillary-driven imbibition in a generated disordered porous medium. The solid particles and sintered necks define the pore space through which the liquid advances, producing a spatially heterogeneous liquid-gas interface. The snapshots show the evolution of the invading liquid at (a) $0$\,ms, (b) $67.5$\,ms, and (c) $300$\,ms.}
    \label{fig:porous_media}
\end{figure*}

The numerical simulations combine a free-surface lattice Boltzmann method
(FSLBM) for the fluid dynamics with a lattice Boltzmann formulation of the
advection-diffusion equation for solute transport. The free-surface method
tracks the evolving liquid-gas interface while neglecting the dynamics of
the gas phase. Solute is released from solid surfaces in contact with the
liquid and is transported by advection and diffusion within the liquid.
The resulting local concentration field determines the fluid viscosity
through Eq.~\eqref{eq:viscosity}, thereby coupling solute transport to the
flow. Details of the numerical implementation are provided in
Appendix~\ref{sec:SI-Methods}.

\subsection{Capillary imbibition}
\label{sec:3-Capillary}

We first simulate the cylindrical capillary geometry considered in
Sec.~\ref{sec:2}, allowing a direct comparison with the analytical
description. As a reference case, we initially consider a pure liquid
imbibing a capillary with non-soluble walls, such that the fluid viscosity
remains constant. The simulation domain consists of a solid cylindrical
tube containing liquid connected to an open boundary, with the remaining
volume occupied by gas at atmospheric pressure. At the liquid boundary,
the density is kept constant, while zero-gradient boundary conditions are
imposed on the axial component of the fluid velocity.
The liquid has a kinematic viscosity
$\nu_0=1\,\mathrm{mm}^2/\mathrm{s}$ and surface tension
$\gamma=1\,\mathrm{mN}/\mathrm{m}$, while the capillary has radius
$r=50\,\mu\mathrm{m}$ and equilibrium contact angle $\theta=60^\circ$.
The conversion between lattice and physical units is detailed in
Appendix~\ref{sec:SI-Units}.

We next introduce dissolution by considering soluble walls. As the liquid advances, solute is released from the wet surface and transported into the liquid, locally modifying its viscosity according to Eq.~\eqref{eq:viscosity}. The relative importance of mass transfer from the wall and diffusion within the liquid is characterized by the Sherwood number,
\begin{equation}
    \mathrm{Sh} = \frac{k_c L}{D},
\end{equation}
where $k_c$ is the mass-transfer coefficient defined in Eq.~\eqref{eq:systemeq} and the characteristic length is again taken as the radius of the tube, $L\equiv r$.

We first test the numerical model in the well-mixed regime considered analytically in Sec.~\ref{sec:2}. Simulations were performed for four mass-transfer coefficients,
$k_c=\{0,10^{-7},10^{-5},10^{-3}\}\,\mathrm{m/s}$, corresponding to
$\mathrm{Sh}=\{0,10^{-6},10^{-4},10^{-2}\}$, while the diffusion coefficient was chosen such that $\mathrm{Pe}=10^{-1}$. The case $k_c=0$ provides the reference system with constant viscosity, whereas increasing $k_c$ produces progressively faster fluid thickening.

Figure~\ref{fig:variable_viscosity} compares the numerical evolution of the height of the meniscus (data points) with the analytical prediction of Eq.~\eqref{eq:varying_viscosity_height} (solid lines). The simulations closely reproduce the analytical curves for the range of dissolution rates considered, including the constant-viscosity limit and the progressive retardation of imbibition as $k_c$ increases. This agreement confirms that the numerical model captures the coupling between dissolution, viscosity evolution, and capillary imbibition in the well-mixed regime.

We next relax the well-mixed approximation by varying the solute diffusion coefficient. As discussed in Sec.~\ref{sec:2}, the assumption of a spatially uniform concentration is expected to hold for $\mathrm{Pe}\ll 1$, when diffusion is fast compared with advective transport. Increasing $\mathrm{Pe}$ therefore allows us to probe the breakdown of the analytical description derived in the well-mixed limit.

Capillary imbibition was then simulated for diffusion coefficients below the well-mixed regime. In these simulations, we considered three distinct diffusion coefficients, $D=\{10^{-7},10^{-6},10^{-5}\}\,\mathrm{m^2/s}$, while keeping the mass-transfer coefficient fixed at $k_c=10^{-5}\,\mathrm{m/s}$. These conditions correspond to $\mathrm{Pe}=\{10,1,10^{-1}\}$, respectively. For $r=50\,\mu\mathrm{m}$, the corresponding Sherwood numbers are $\mathrm{Sh}=\{10^{-2}, 10^{-3}, 10^{-4}\}$. The results are shown in Fig.~\ref{fig:diffusion}, together with the analytical prediction of Eq.~\eqref{eq:varying_viscosity_height}, corresponding to the well-mixed limit.

As the diffusion coefficient increases, concentration gradients within the liquid are progressively suppressed and the numerical results approach the
well-mixed analytical prediction. Conversely, for slower diffusion~(${\mathrm{Pe}\gtrsim 1}$), the solute concentration becomes spatially heterogeneous, producing a corresponding heterogeneous viscosity field. In this regime, the retardation of the advancing meniscus is weaker than predicted by the well-mixed theory. Thus, although dissolution increases
the fluid viscosity locally, its effect on the imbibition dynamics depends on how efficiently the dissolved material is redistributed throughout the
liquid column.

\subsection{Imbibition in disordered porous media}
\label{sec:3-Porous_media}
\begin{figure}[b]
    \includegraphics[width=\linewidth]{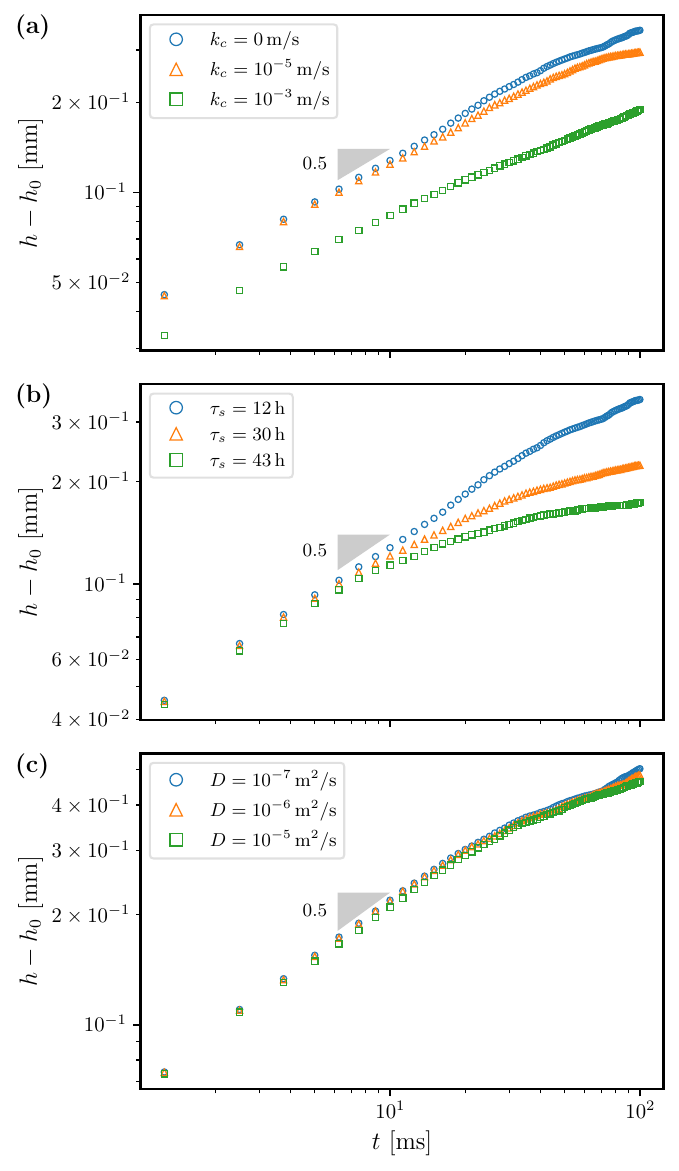}
    
    \caption{Mean interface height during capillary-driven imbibition in disordered sintered porous media, averaged over $10$ independently generated structures. (a) Effect of the mass-transfer coefficient $k_c$: increasing the dissolution rate enhances fluid thickening and slows imbibition. (b) Effect of the sintering time $\tau_s$ in the absence of dissolution: increasing $\tau_s$ promotes neck growth and restricts the pore space, reducing the advance of the liquid. (c) Effect of the diffusion coefficient $D$ at $\tau_s=12\,\mathrm{h}$ and $k_c=10^{-4}\,\mathrm{m/s}$: faster diffusion produces a modest additional retardation by redistributing the dissolved material more uniformly throughout the invading liquid.
    \label{fig:pm_results}}
\end{figure}
We now extend the analysis from a single cylindrical capillary to disordered porous media, again neglecting gravity. In this case, the invading liquid
advances through an interconnected network of pores with different sizes and geometries, resulting in a spatially heterogeneous interface and more complex
imbibition dynamics. Our aim is to determine how the dissolution-driven fluid thickening identified in the capillary geometry affects imbibition in a disordered pore space, and how this effect compares with variations in the geometry of the porous medium.

We consider sintered granular media, which provide disordered pore structures relevant to powder-based materials, including those encountered in food processing. The porous media were generated by first relaxing assemblies of spherical particles with different sizes using a discrete-element method. The resulting packings were then subjected to a controlled sintering process~\cite{Braz2022}, during which neighboring particles developed solid necks at their contact points. The resulting structures were mapped onto the LBM grid, with the particles and their connecting necks defining the soluble solid phase. The liquid and gas phases were initialized in the remaining pore space following the same procedure used for the capillary simulations. An example of the resulting porous medium and the evolution of the invading liquid is shown in
Fig.~\ref{fig:porous_media}.

We generated $10$ independent porous structures in cubic domains with side length $2.56\,\mathrm{mm}$. The particle radii were drawn from a Weibull distribution with shape parameter $a=1.5$ and scale parameter $b=128\,\mu\mathrm{m}$~\cite{Braz2022}. For each structure, we simulated imbibition while varying the mass-transfer coefficient $k_c$ and the sintering time $\tau_s$. Increasing $\tau_s$ amplifies neck growth and thereby modifies the geometry of the pore space.  Because the liquid-gas interface is highly irregular, its position was characterized by the mean height of the interface cells. The invasion dynamics also exhibits substantial sample-to-sample fluctuations associated with the disordered pore geometry. We therefore average the interface height over the $10$ independently generated structures.

Figure~\ref{fig:pm_results} shows how the mean interface height depends on dissolution, pore geometry, and solute diffusion. We first vary the mass-transfer coefficient $k_c$ while keeping the porous structure fixed ($\tau_s=12\,$h and $D=10^{-5}\,$m$^2$/s). As in the capillary geometry, increasing $k_c$ accelerates solute uptake and the associated increase in fluid viscosity, leading to slower imbibition. This effect becomes particularly pronounced at the largest $k_c$, for which the viscosity increases rapidly and substantially reduces the penetration of the liquid into the porous
medium (see Fig.~\ref{fig:pm_results}a).

To separate this rheological effect from that of the pore geometry, we next vary the sintering time $\tau_s$ in the absence of dissolution ($k_c=0$), so that the liquid viscosity remains constant. Increasing $\tau_s$ promotes neck growth between neighboring particles and reduces the pore space available to the invading liquid. For the shortest sintering time, neck growth is negligible and the structure remains close to a packing of spherical particles. At longer sintering times, the growing necks progressively narrow the pore throats and hinder liquid transport. Accordingly, Fig.~\ref{fig:pm_results}b shows a substantial reduction of the imbibition rate with increasing $\tau_s$. This demonstrates that changes in pore geometry can produce a retardation comparable to that caused by dissolution-driven fluid thickening, although through a distinct mechanism.

Finally, we examine the effect of solute diffusion for structures with $\tau_s=12\,\mathrm{h}$ and $k_c=10^{-4}\,\mathrm{m/s}$. Consistent with the capillary results, increasing $D$ leads to a slightly lower position of the liquid front (see Fig.~\ref{fig:pm_results}c). Faster diffusion redistributes the dissolved material more efficiently throughout the invading liquid, making the viscosity field more homogeneous and increasing the overall retardation relative to the case of slower diffusion. Within the range investigated, however, the effect of varying $D$ is considerably weaker than that produced by changes in either the dissolution rate or the pore geometry.

\section{Conclusions}\label{sec:4}
We have investigated how dissolution-driven changes in fluid viscosity modify capillary imbibition. In the well-mixed limit (${\mathrm{Pe} < 1}$), we derived a generalized Washburn's law showing that the effect of an evolving viscosity can be incorporated through a rescaled time determined by the accumulated inverse viscosity, $\tilde{t}=\int_0^t[\eta_0/\eta(t')]\,\mathrm{d}t'$. This result is independent of the specific concentration-viscosity relation and shows that, as long as the fluid remains spatially well mixed, the classical Washburn dynamics are recovered when expressed in terms of this effective time. For the specific dissolution and viscosity model considered here, the theory predicts a crossover between Washburn-like regimes governed by the initial and asymptotic viscosities.

Numerical simulations of capillary imbibition confirm this analytical description over a range of dissolution rates. However, when the well-mixed condition is relaxed (${\mathrm{Pe}\gtrsim 1}$), concentration gradients develop within the invading liquid and produce a spatially heterogeneous viscosity field. In this regime, the retardation of the advancing meniscus is weaker than predicted by the well-mixed theory. Increasing the diffusion coefficient progressively homogenizes the concentration field and drives the dynamics toward the analytical well-mixed limit. Thus, the effect of dissolution on imbibition is controlled not only by the resulting increase in viscosity, but also by how the dissolved material is distributed throughout the invading liquid.

Extending the analysis to disordered porous media shows that the same dissolution-driven mechanism remains important in geometrically heterogeneous pore spaces. Increasing the dissolution rate enhances fluid thickening and reduces the advance of the liquid front, while slower diffusion weakens this retardation by maintaining a more heterogeneous concentration and viscosity field. Independently, increasing the degree of sintering modifies the pore geometry by promoting neck growth and narrowing pore throats, producing a substantial reduction in imbibition even when the fluid viscosity remains constant. Within the range of parameters investigated here, variations in the dissolution rate and pore geometry have a stronger effect on imbibition than variations in solute diffusion.

These results provide a quantitative extension of the classical Washburn picture to situations in which interactions with the solid progressively modify the rheology of the invading liquid. More generally, they highlight a feedback mechanism complementary to the structural evolution commonly considered in reactive and evolving porous media: fluid-solid interactions can strongly modify transport even when the geometry of the porous matrix remains fixed, simply by changing the properties of the invading fluid. The framework developed here provides a basis for investigating this coupling in more complex systems, including deformable or evolving porous structures, reactive transport, filtration, and porous materials relevant to food processing.

\section*{Acknowledgements}
We acknowledge financial support from the Portuguese Foundation for Science and Technology (FCT) under the contracts no. UID/00618/2025 (https://doi.org/10.54499/UID/00618/2025), UID/PRR2/00618/2025 \\(https://doi.org/10.54499/UID/PRR2/00618/2025), UID/PRR/00618/2025 \\(https://doi.org/10.54499/UID/PRR/00618/2025), 2023.10412.CPCA.A2 (DOI 10.54499/2023.10412.CPCA.A2), and FCT/Mobility/1348751812/2024-25. R.C. acknowledges financial support from FAPERJ – Fundação Carlos Chagas Filho de Amparo à Pesquisa do Estado do Rio de Janeiro (Process No. SEI-260003/020878/2025) and from CNPq – Conselho Nacional de Desenvolvimento Científico e Tecnológico (Process No. 317149/2025-3). A. F. V. M. thanks support from the European Union's Horizon Europe research and innovation program under Grant Agreement No. 101203506, Marie Sklodowska-Curie Action Postdoctoral Fellowship, project IonFlowElast.

\appendix
\section{Numerical methods} \label{sec:SI-Methods}
\subsection{Lattice Boltzmann method} \label{sec:SI-LBM}
Lattice Boltzmann methods~\cite{Krger2017, 10.1093/oso/9780199592357.001.0001} are a class of computational fluid dynamics methods based on kinetic theory that determine the macroscopic flow of a liquid by computing the evolution of particle distribution functions on a discretized lattice.

In this work, the D3Q19 lattice was employed, consisting of a three-dimensional grid with a set of 19 velocity vectors $\bm{c}_i, i \in \{0, 1, ..., 18\}$. In each cell of the grid, there are 19 different distributions $f_i$, corresponding to each velocity vector. The discretized Boltzmann equation gives the time evolution of these distributions as a function of a collision operator,
\begin{equation}
    f_i(\textbf{x} + \bm{c}_i \Delta t, t + \Delta t) = f_i(\textbf{x}, t) + \left( \pdv{f}{t} \right)_\text{coll}.
\end{equation}

The collision operator is efficiently approximated using the single-relaxation-time BGK operator~\cite{bgk1954collision},
\begin{equation}
    \left( \pdv{f}{t} \right)_\text{coll} = - \frac{f_i(\textbf{x}, t) - f_i^\text{eq}(\rho_G, \bm u)}{\tau_f} \, ,
\end{equation}
which assumes that the particle distribution functions $f_i$ converge exponentially to an equilibrium distribution $f^{eq}$ within a relaxation time $\tau_f$. The equilibrium distribution functions are defined by the following expression:
\begin{equation}
    f^{\text{eq}}_i = w_i \rho \left[ 1 + \frac{\mathbf{c}_i \cdot \mathbf{u}}{c_s^2} - \frac{u^2}{2 c_s^2} + \frac{(\mathbf{c}_i \cdot \mathbf{u})^2}{2 c_s^4} \right],
    \label{eq:equilibrium}
\end{equation}
where $w_i$ are the weights related to each velocity vector and are characteristic of the chosen velocity discretization. The density $\rho$ and the macroscopic velocity $\textbf{u}$ of the fluid are obtained from the zeroth and first moments of the distribution function,
\begin{equation}
    \rho = \sum_i f_i, \quad \textbf{u} = \frac{1}{\rho} \sum_i \bm{c}_i f_i \, .
\end{equation}

The kinematic viscosity is a function of the relaxation constant $\tau_f$ and the speed of sound in the lattice $c_s$,
\begin{equation}
    \nu = c_s^2 \left( \tau_f - \frac{\Delta t}{2} \right) \, .
    \label{eq::def.viscosity}
\end{equation}
We assume that variations in solute concentration do not significantly affect the fluid density, such that $\rho$ can be treated as constant. Since $\eta=\rho\nu$, the dynamic and kinematic viscosities then have the same relative dependence on concentration.

\subsection{Free-surface lattice Boltzmann method} \label{sec:SI-FSLBM}
The FSLBM combines a lattice Boltzmann description of the liquid dynamics with a volume-of-fluid (VOF) approach to track the liquid--gas interface. The implementation used in this work is mainly based on the formulations described in Refs.~\cite{THORIMBERT2021101338,PhysRevE.93.043302, 10.1063/5.0131159,Korner2005}. The dynamics of the gas phase are neglected, since their influence on the liquid is represented through a free-surface boundary condition at the liquid-gas interface. This boundary condition imposes zero shear stress and a normal stress balance accounting for the local gas pressure and the Laplace pressure.

The free-surface is tracked using the volume-of-fluid (VOF) method. For this reason, this free-surface model can also be referred to as LB-VOF. In addition to the common LBM variables, two new variables are tracked in each cell: the fill level $\varphi$ and the mass $m$. These two variables are related through the local liquid density,
\begin{equation}
    \varphi = \frac{m}{\rho}.
\end{equation}

This model distinguishes three different types of cells according to their fill level. Gas cells are defined by $\varphi = 0$, liquid cells by $\varphi = 1$, and interface cells by ${0 < \varphi < 1}$. Gas cells are ignored because their dynamics do not influence the rest of the system, liquid cells are simulated using the common LBM approach, and the interface cells satisfy the free-surface condition that accounts for the interaction with the gas phase. This condition is implemented by reconstructing the missing distribution functions that would have originated from gas cells. These missing populations are calculated using the following expression:
\begin{align}
    f_I^{out}(\bm x + \bm{c}_i \Delta t, t) = &f_i^{eq}(\rho_G, \bm u) + f_I^{eq}(\rho_G, \bm u) \nonumber \\ &- f_i^{out}(\bm x, t) \, ,
\end{align}
where $f_I^{out}$ is the missing population after the collision step and before the streaming step. The $I$ denotes the opposite direction of $i$, $\rho_G$ is the gas density at the interface, and $\bm u$ is the macroscopic velocity at the interface cell.

In simpler models, the gas density at the interface can simply be set to unity. The gas will be under atmospheric pressure and the liquid will have no surface tension. By contrast, in this model, the gas pressure incorporates the volume pressure at the interface $p_V(t)$, resulting from volume constrictions, and the local Laplace pressure $p_L(\bm x, t)$, which arises from the surface tension of the liquid,
\begin{equation}
    p_G(\bm x, t) = p_V(t) + p_L(\bm x, t) \, .
\end{equation}
The density and pressure of the gas at the interface are related by the equation of state,
\begin{equation}
    \rho_G = \frac{p_G}{c_s^2}.
\end{equation}

The volume pressure can be easily determined after labeling each distinct gas bubble and updating their volume after each iteration,
\begin{equation}
    p(t) = p(0) \frac{V(0)}{V(t)} \, .
\end{equation}
Alternatively, it can be set to constant when the gas is at atmospheric pressure, which was the case in all systems studied here. The algorithm responsible for tracking the volume of gas bubbles is also capable of computing the coalescing and splitting of bubbles~\cite{Pohl2008, Siko2012, ANDERL2014331}.

The Laplace pressure is computed from
\begin{equation}
    p_L(\bm x, t) = 2 \gamma \kappa(\bm x, t) \, ,
\end{equation}
where $\gamma$ is the surface tension coefficient and $\kappa$ is the local curvature of the interface. The interface normal $\bm n$ is computed from the gradient of the fill level function, $\bm n(\bm x, t) = \nabla \varphi$, and a detailed description of its implementation can be found in~\cite{PhysRevE.93.043302}. Then, the local curvature of the interface is estimated from the divergence of the normalized interface normal, $\kappa = -\nabla \cdot \hat{\bm n}$, and the Laplace pressure is determined. The gradient and the divergence were computed using their corresponding lattice differential operators \cite{Ramadugu_2013}. The wetting properties of the liquid are set by the equilibrium contact angle~$\theta$, which determines the direction of the unit normal $\hat{\bm n}$ at the solid nodes that neighbor the liquid interface~\cite{PhysRevE.93.043302}.

After this step, the distribution functions can be streamed as is usually done in the LBM and then the mass of each cell is updated based on the evolution of the populations,
\begin{align}
    &m(\bm x, t + \Delta t) = m(\bm x, t) + \sum_{i=0}^Q \Delta m_i(\bm x, t + \Delta t),\\
    &\Delta m_i(\bm x, t + \Delta t) = \frac{\varphi(\bm x) + \varphi(\bm x + \bm{c}_i \Delta t)}{2} \Delta f_i \, ,
\end{align}
where changes in mass, $\Delta m_i$, are a function of changes in the distributions $\Delta f_i$,
\begin{equation}
    \Delta f_i = f_I(\bm x, t) - f_i(\bm x + \bm{c}_i \Delta t, t) \, .
\end{equation}

Finally, the fill level in each cell is updated and the interface cells are converted to liquid cells if ${\varphi > 1 + \epsilon}$ or gas cells if ${\varphi < -\epsilon}$. The threshold $\epsilon$ prevents quick oscillatory conversions and is set to $10^{-2}$. After the interface cells are converted, extra conversions need to be performed to preserve the continuity of the interface surface and to prevent liquid and gas cells from being direct neighbors. For conversions of interface cells to gas or liquid cells, the fill level will be directly modified to ${\phi = 0}$ or ${\phi = 1}$, respectively. Generally, this results in an excess of mass in the modified cells that has to be redistributed to neighboring interface cells.

\subsection{Advection-diffusion model} \label{sec:SI-Dissolution}
The movement of a dissolved solute through a porous medium is governed by advection (solute particles are carried along the flowing fluid) and diffusion (driven by Brownian motion). Together with any dissolution processes, these mechanisms are described by the advection–diffusion equation,
\begin{equation}
    \pdv{C}{t} + \div(\mathbf{u} \, C) = \div(D\,\grad C) + S \, ,
\end{equation}
where $C$ is the solute concentration, $\bm u$ is the macroscopic velocity of the liquid, $D$ is the diffusion coefficient, and $S$ is the dissolution rate.

This equation is numerically integrated using a second lattice Boltzmann method, following the formulation described by Matias et al.~\cite{Matias10.1063/5.0176273}, and coupled to the FSLBM. The advection-diffusion solver introduces a new set of distributions $g_i$ on the same D3Q19 lattice, whose zeroth moment gives the local solute concentration.

The advection and diffusion of the solute are modeled by evolving the distributions as presented in Section \ref{sec:SI-LBM}. To introduce the dissolution of the solute, a source term is added to the collision step,
\begin{align}
    g_i(\textbf{x} + \bm{c}_i \Delta t, t + \Delta t) = &g_i(\textbf{x}, t) + \left( \pdv{g}{t} \right)_\text{coll} \nonumber \\
    &+ \left(1 - \frac{1}{2 \tau_g}\right) w_i S \, ,
\end{align}
where $\tau_g$ is the relaxation time parameter, which is related to the diffusion coefficient $D = c_s^2 \left( \tau_g - \Delta t/2 \right)$. The dissolution rate $S$ is directly proportional to the surface area of the solute-liquid interface and to the mass transfer rate coefficient $k_c$. It is also a function of the concentration at the surface $C_s$, the concentration in the liquid bulk $C_l$, and the liquid volume $v_l$ in the cell where dissolution occurs~\cite{Matias10.1063/5.0176273},
\begin{equation}
    S = k_c (C_s - C_l) \frac{A}{v_l} \, .
    \label{eq:source_term}
\end{equation}

In this work, we considered that the solute concentration in the liquid bulk is always much lower than the solute concentration at the solid surface ($C_l \ll C_s$), and that the grains maintain their initial structure, not suffering from swelling or erosion. The solute concentration at the solid surface was considered to be of the order of the water density ($C_s \approx 1000\,$kg/m$^3$). The dissolution rate is therefore approximated by $S \approx k_c C_s A / v_l$, where both $C_s$ and $v_l$ are taken as $1$ in lattice units. Hence, the mass transfer rate coefficient is the parameter that sets the dissolution rate, as the surface area of the solute-liquid interface is constant. The surface area is estimated by the components of the vector normal to the surface ~\cite{MATIAS2021101360, Matias10.1063/5.0176273},
\begin{equation}
    A = \frac{\sqrt{1+n_2^2}\sqrt{1+n_3^2}}{(1+n_2)(1+n_3)} \, ,
\end{equation}
where $n_i$ are the normal vector components scaled in such a way that the smallest non-zero component is set to unity.

The equilibrium distribution function $g^{eq}$ has an expression similar to that of Eq.~\eqref{eq:equilibrium}, where the first moment now represents the concentration of solute $C$ and the second moment is still given by the macroscopic velocity of the liquid, which is computed by the FSLBM.

The advection-diffusion solver is coupled to the fluid dynamics through the local velocity and concentration fields. The velocity field computed by the FSLBM determines the advective transport of solute, while the resulting concentration field is used to locally update the kinematic viscosity according to Eq.~\eqref{eq:viscosity}. Since the kinematic viscosity is related to the fluid relaxation time $\tau_f$ through Eq.~\eqref{eq::def.viscosity}, the local value of $\tau_f$ is updated accordingly. The coupling therefore acts in both directions: the fluid velocity determines solute transport, while solute uptake modifies the local viscosity and, consequently, the fluid dynamics.

\section{Washburn limits}\label{sec:SI-Washburn}
The analytical solution for the capillary rise of a liquid with varying viscosity specified by equation \eqref{eq:viscosity} is given by equation \eqref{eq:varying_viscosity_height},
\begin{align}
    h(t) = &\frac{1}{\beta + 1} \left\{ a (\beta + 1) t + h_0^2 (\beta+1)^2 \right. \nonumber \\ &+ \left. a \beta d \log \left[1 + \frac{(\beta + 1)}{d} t \right]\right\}^{1/2} \, ,
\end{align}
where $a = \gamma r \cos \theta / 2 \eta_0$ and $d = \delta r / 2 k_c C_s$.

In the limit $k_c \ll \delta \, r / C_s$, $d$ tends to infinity, and the logarithmic term is approximated to first order,
\begin{align}
    \lim_{d \to \infty} h(t) \approx &\frac{1}{\beta + 1} \biggl\{ a (\beta + 1) t + h_0^2 (\beta+1)^2 \bigl. \nonumber \\ &+ \bigr. a \beta (\beta + 1) t \biggr\}^{1/2} \nonumber \\
    = &\frac{1}{\beta + 1} \sqrt{a (\beta + 1)^2 t + h_0^2 (\beta+1)^2} \\
    = &\sqrt{a t + h_0^2} \nonumber \\
    = &\sqrt{\frac{\gamma r \cos \theta}{2 \eta_0} t + h_0^2} \; . \nonumber
\end{align}
Since the viscosity of the liquid remains equal to the intrinsic viscosity $\eta_0$, the height of the meniscus follows Washburn's law.

In the limit $k_c \gg \delta \, r / C_s$, $d$ tends to zero, and the logarithmic term is neglected,
\begin{align}
    \lim_{d \to 0} h(t) \approx &\frac{1}{\beta + 1} \sqrt{a (\beta + 1) t + h_0^2 (\beta+1)^2} \nonumber \\
    = &\sqrt{\frac{a}{\beta + 1} t + h_0^2} \\
    = &\sqrt{\frac{\gamma r \cos \theta}{2 \eta_0 (1 + \beta)} t + h_0^2} \, . \nonumber
\end{align}
In this case, the viscosity quickly reaches its maximum value $\eta_{\mathrm{sat}} = \eta_0 (1 + \beta)$, and the height of the meniscus follows Washburn's law corresponding to a liquid with viscosity $\eta_{\mathrm{sat}}$.

\section{Conversion of units}\label{sec:SI-Units}
All input and output values of the simulation are initially given in lattice units (l.u.), but these values are later converted to standard units using conversion factors for length ($C_l$), time ($C_t$), and mass ($C_m$).

The size of the solute grains determined the length conversion factor. Since these particles have diameters of around hundreds of micrometers~\cite{Dupas2017}, the length factor was chosen as $C_l = 10 \, \mu$m. The time conversion factor $C_t = 1 \, \mu$s, was obtained by equating the kinematic viscosity of the liquid in the model, $\nu = 0.01 \,$l.u., to the physical kinematic viscosity of water, $\nu_\text{water} = 1 \,$mm$^2$/s. Analogously, by equating the reference density in the model, $\rho = 1 \,$l.u., to the density of water, $\rho_\text{water} = 1000 \,$kg/m$^3$, the mass conversion factor is given by $C_m = 10^{-12} \,$kg.

Although these conversion factors were specifically chosen for the imbibition process in porous media, the same values were used in the capillary systems.

\bibliographystyle{apsrev4-2}
\tiny
\bibliography{paper}

\end{document}